\documentclass[11pt]{article}
\usepackage{amsthm}
\usepackage{amsmath}
\usepackage{amssymb}
\usepackage{makeidx}
\usepackage{latexsym}
\usepackage{epsf}
\usepackage{epsfig}

\newtheorem{theorem}{Theorem}             
\newtheorem{proposition}{Proposition}        
\newtheorem{corollary}{Corollary}           
\newtheorem{lemma}{Lemma}               
\newtheorem{rul}{Rule}               
\newtheorem{remark}{Remark}              
\newtheorem{definition}{Definition}   

\def\NP{\mbox{\sc{NP}\/}}

\def\DTIME{\mbox{\sc{DTIME}\/}}

\def\None{N_1} 
\def\Ntwo{N_2} 
\def\Nthree{N_3} 

\begingroup
\catcode`@=11\relax%
\catcode`{=12\relax\catcode`}=12\relax%
\catcode`(=1\relax \catcode`)=2\relax%
\gdef\includeversion#1(%
  \expandafter\gdef\csname #1\endcsname%
    ()%
  \expandafter\gdef\csname end#1\endcsname%
    ()%
)%
\gdef\excludeversion#1(%
  \expandafter\gdef\csname #1\endcsname%
    (\@bsphack\catcode`{=12\relax\catcode`}=12\relax\csname #1@NOTE\endcsname)%
  \long\expandafter\gdef\csname #1@NOTE\endcsname ##1\end{#1}%
    (\csname #1END@NOTE\endcsname)%
  \expandafter\gdef\csname #1END@NOTE\endcsname%
    (\@esphack\end(#1))%
)%
\endgroup
\excludeversion{comment}

\begin{document}
\title{
Polynomial Time Data Reduction for {\sc Dominating Set}\footnote{%
An extended abstract of this work entiteled
``Efficient Data Reduction for {\sc Dominating Set}: A Linear Problem Kernel for the Planar Case''
appeared in the {\em Proceedings
of the 8th Scandinavian Workshop on Algorithm Theory (SWAT 2002)},
Lecture Notes in Computer Science (LNCS) 2368, pages 150--159, Springer-Verlag 2002.}}
\author{Jochen Alber\thanks{%
Contact author. Wilhelm-Schickard-Institut f\"ur Informatik,
Universit\"at T\"ubingen, Sand 13, D-72076~T\"ubingen,
Fed.\ Rep.\ of Germany.
Email: alber@informatik.uni-tuebingen.de.
           Work supported by the Deutsche Forschungsgemeinschaft (DFG), research project PEAL
           (Parameterized complexity and Exact ALgorithms), NI 369/1-1,1-2.}
\and Michael~R.~Fellows\thanks{%
Department of Computer Science and Software Engineering,
University of Newcastle, University Drive, Callaghan 2308, Australia.
Email: mfellows@cs.newcastle.edu.au.
}
\and 
Rolf Niedermeier\thanks{%
Wilhelm-Schickard-Institut$\,$f\"ur$\,$Informatik,$\,$%
Universit\"at$\,$T\"ubingen,$\,$Sand$\,$13,$\,$D-72076$\,\,$T\"ubingen,$\,$%
Fed.$\,$Rep.$\,$of Germa\-ny.
Email: niedermr@informatik.uni-tuebingen.de.
}}

%
%
\date{}
\maketitle 

\vspace*{-0.5cm}

\vspace*{-0.5cm}
\begin{abstract}
\noindent Dealing with the NP-complete 
{\sc Dominating Set} problem on undirected graphs, we demonstrate the 
power of data reduction by preprocessing from
a theoretical as well as a practical side.
In particular, we prove that {\sc Dominating Set}
restricted to planar graphs has a so-called problem kernel 
of linear size, achieved by two simple and easy to
implement reduction rules. 
Moreover, having implemented our reduction rules,
first experiments 
indicate the impressive practical potential of these rules. 
Thus, this work seems to open up a new and prospective way how to cope with
one of the most important problems in graph theory and combinatorial optimization.
\end{abstract}

\section{Introduction}
{\bf Motivation.}
A core tool for practically solving NP-hard problems 
is data reduction through preprocessing.
Weihe~\cite{Wei98,Wei00} gave a striking example when dealing
with the NP-complete {\sc Red/Blue Dominating Set} problem
appearing in context of the European railroad network. In a preprocessing phase,
he applied two simple data reduction rules again and again
until no further application was possible. The impressive
result of his empirical study was that each of his real-world instances
was broken into very small pieces such that for each of these a simple
brute-force approach was sufficient to solve the hard problems
efficiently and optimally.
In this work, we present a new and stronger example for data reduction
through preprocessing, namely for the NP-complete 
{\sc Dominating Set} problem, a 
core problem of combinatorial
optimization and graph theory.
According to a~1998 survey~\cite[Chapter 12]{HHS98},
more than 200~research papers and more than 30~PhD theses
investigate the algorithmic complexity of domination and related
problems~\cite{Tel94}.
Moreover, domination problems occur in numerous practical settings,
ranging from strategic decisions such as locating radar stations or
emergency services through computational biology to voting systems
(see \cite{HHS98,HHS98b,Rob78} for a survey).
By way of contrast to the aforementioned example given by Weihe, 
however, our preprocessing is, on the one hand, more involved to 
develop, and, on the other hand, it does not only prove its strength 
through experimentation but, in first place, by theoretically sound 
means. Thus, we come up with a practically promising as well 
as theoretically appealing result for computing the domination number 
of a graph, one of the so far few positive news for this 
important problem.

\medskip
\noindent 
{\bf Problem definition and status.}
A {\em $k$-dominating set\/}~$D$ of an undirected graph~$G$
is a set of $k$~vertices of~$G$ such that each of the rest
of the vertices has at least one neighbor in~$D$. The minimum~$k$
such that $G$~has a $k$-dominating set is called the
{\em domination number\/} of~$G$, denoted by~$\gamma (G)$.
The {\sc Dominating Set} problem is to decide, given a
graph~$G=(V,E)$ and a positive integer~$k$, whether
$\gamma (G)\leq k$.
Due  to its NP-completeness and its practical importance,
{\sc Dominating Set} has been subject to intensive studies that were
concerned with coping strategies to attack its intractability.
Among these coping strategies, we find approximation 
algorithms and (exact) fixed-parameter algorithms.
As to approximation results, it is known that 
{\sc Dominating Set} is polynomial time approximable with
factor $1+\log |V|$ since the problem is a special case of 
the {\sc Minimum Set Cover} problem~\cite{Joh74}.
On the negative side, however, it is known not 
to be approximable within $(1-\epsilon )\ln |V|$ for any $\epsilon >0$
unless $\NP\subseteq \DTIME(n^{\log\log n})$~\cite{Fei98}.
When restricted to planar graphs, where it still remains 
NP-complete~\cite{GJ79}, however, a polynomial time approximation
scheme (PTAS) is stated~\cite{Bak94}.\footnote{In~\cite{Bak94},
only the conceptually much simpler {\sc Independent Set} problem 
is described in detail.}
There are numerous approximation results for further 
special instances of {\sc Dominating Set},
cf.~\cite{ACGKMP99}.
As to fixed-parameter results, the central question is whether the problem
is optimally solvable in time 
$f(k)\cdot n^{O(1)}$, where $f(k)$ may be an exponentially fast (or worse)
growing function in the {\em parameter\/}~$k$ only and 
$n$~is the number of graph vertices. 
Unfortunately, also here the situation seems hopeless---the problem 
is known to be W[2]-complete~\cite{DF92,DF99}
which implies fixed-parameter intractability 
unless very unlikely collapses of parameterized complexity classes occur
(see~\cite{DF99} for details).
Again, restricting the problem to planar graphs improves the situation.
Then, {\sc Dominating Set} is known to be solvable in time
$O(c^{\sqrt{k}} \cdot n)$ for $c\leq 
4^{6\sqrt{34}}$~\cite{ABFN00}\footnote{Note that in the SWAT 2000 conference version
of~\cite{ABFN00}, an exponential
base~$c=3^{6\sqrt{34}}$ is stated, caused by a misinterpretation of previous results.
The correct worst-case
upper bound reads $c=4^{6\sqrt{34}}$.}
and, alternatively, solvable in time $O(8^k \cdot n)$~\cite{AFFFNRS01}.
As to fixed-parameter complexity, it was open whether 
{\sc Dominating Set} on planar graphs possesses a so-called 
problem kernel of linear size which we answer 
affirmatively here.

\medskip
\noindent
{\bf Results.}
We provide positive news on the algorithmic tractability 
of {\sc Dominating Set} through preprocessing.
The heart of our results are two relatively simple and easy to implement 
``reduction rules'' for {\sc Dominating Set}.
These rules are based on considering local structures 
within the graph. They produce a reduced graph such that the original graph
has a dominating set of size at most~$k$ iff 
the reduced graph has a dominating set of size at most~$k'$ for
some $k'\leq k$.
The point here is that the reduced graph, as a rule, is much smaller 
than the original graph and, thus, $k'$~is 
significantly smaller than~$k$ because the reduction process usually 
determines several vertices that are part of an optimal dominating set.
In this way, these two reduction rules provide an efficient data 
reduction through polynomial time preprocessing.
In the case of planar graphs, we actually can prove that the reduced 
graph consists of at most $335k$~vertices (which
is completely independent of the size of the original graph).
In fixed-parameter complexity terms, this means that 
{\sc Dominating Set} on planar graphs possesses a linear size problem
kernel.
Note, however, that our main concern in analyzing
the multiplicative constant~$335$ was conceptual simplicity
for which we deliberately sacrificed the aim to further
lower it by way of refined analysis (without changing the
reduction rules).
Finally, experimental studies underpin the big potential 
of the presented reduction rules, leading to graph size reductions of more
than 90~percent when experimenting with random graphs.
Hence, we anticipate that every future algorithm for
{\sc Dominating Set}, whether approximation, fixed-parameter,
or purely heuristic, always should employ the data reduction method 
proposed here.

\medskip
\noindent
{\bf Relation to previous work.}
Our data reduction still allows to solve the problem exactly,
not only approximately. It is, thus, always possible to incorporate 
our reduction rules in any kind of approximation algorithm 
for {\sc Dominating Set} without deteriorating its approximation 
factor. Baker's PTAS result for 
{\sc Dominating Set} on planar graphs~\cite{Bak94} probably 
has much less applicability than the result presented here.
This is due to the fact that, as a rule, PTAS algorithms rarely are
efficient enough in order to be of practical use.
Also, our, data reduction algorithm is conceptually much simpler
and, as a preprocessing method, seems to combine with 
{\em any\/} kind of algorithm working afterwards on the then reduced graph.
Concerning the parameterized complexity of 
{\sc Dominating Set} on planar graphs, we have the following consequences 
of our result.
First, on the structural side, combining our linear problem kernel
with the graph separator approach presented in~\cite{AFN01b}
immediately results in an $O(c^{\sqrt{k}}\cdot k + n^{O(1)})$
{\sc Dominating Set} algorithm on planar graphs (for some
constant~$c$). Also, the linear problem kernel directly proves the
so-called
``Layerwise Separation Property''~\cite{AFN01a} for
{\sc Dominating Set} on planar graphs, again implying an
$O(c^{\sqrt{k}}\cdot k + n^{O(1)})$
algorithm.
Second,
the linear problem kernel improves the time $O(8^k\cdot n)$ search tree
algorithm from~\cite{AFFFNRS01} to an $O(8^k k + n^{O(1)})$ algorithm.
We are aware of only
one further result that provides a {\em provable\/} data 
reduction by preprocessing in our sense,
namely the Nemhauser-Trotter theorem for 
{\sc Vertex Cover}~\cite{NT75,BE85}. 
Their polynomial time preprocessing employs a maximum matching 
algorithm for bipartite graphs and provides a reduced graph where at 
least half of the vertices have to be part of an optimal 
vertex cover set (also see~\cite{CKJ01} for details and its
implication of a size~$2k$ problem kernel).
Note, however, that from an algorithmic and combinatorial
point of view, {\sc Vertex Cover} seems to be a much less elusive 
problem than {\sc Dominating Set} is.

\medskip
\noindent 
{\bf Structure of the paper.}
We start with our two reduction rules based on the neighborhood structure 
of a single vertex and a pair of vertices, respectively.
Here, we also analyze the worst-case time complexity 
of these reduction rules for planar as well as for general graphs.
Afterwards, in the technically most demanding part, we prove that 
for planar graphs our reduction rules 
{\em always\/} deliver a reduced graph of size 
$O(\gamma (G))$.
Finally, we discuss some first experimental findings
and give some conclusions and challenges for future work.

\section{The Reduction Rules}
\label{rules}

We present two reduction rules
for {\sc Dominating Set}. Both reduction 
rules are based on the same principle: We explore 
local structures of the graph and try to replace them 
by simpler structures. 
For the first reduction rule, the local 
structure will be the neighborhood of a single vertex.
For the second reduction rule, we will deal 
with the union of the neighborhoods of a pair of vertices.
\subsection{The Neighborhood of a Single Vertex}

Consider a vertex $v\in V$ of the given graph~$G=(V,E)$. 
We partition the vertices of the neighborhood~$N(v)$ of~$v$
into three different sets $\None(v)$, $\Ntwo(v)$, and $\Nthree(v)$
depending on what neighborhood
structure these vertices have. More precisely,  setting $N[v] := N(v) \cup \{ v\}$, we 
define
\begin{eqnarray*}
\None(v) & := & \{ u \in N(v) \,:\,  N(u) \setminus N[v] \not= \emptyset \},\footnotemark \\
\Ntwo(v) & := & \{ u \in N(v)\setminus \None(v) \,:\, N(u) \cap \None(v) \not= \emptyset \}, \\
\Nthree(v) & := & N(v) \setminus (\None(v) \cup \Ntwo(v)).
\end{eqnarray*}

\footnotetext{For two sets~$X$,$Y$, where $Y$ is not necessarily a subset of~$X$, we use the convention 
that $X\setminus Y:=\{x \in X: x \notin Y\}$.}

\begin{figure}[t]
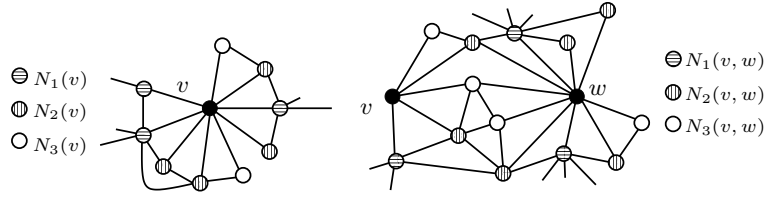

\begin{center}
\begin{picture}(0,0)%
\epsfig{file=rule1.pstex}%
\end{picture}%
\setlength{\unitlength}{1450sp}%
\begingroup\makeatletter\ifx\SetFigFont\undefined%
\gdef\SetFigFont#1#2#3#4#5{%
  \reset@font\fontsize{#1}{#2pt}%
  \fontfamily{#3}\fontseries{#4}\fontshape{#5}%
  \selectfont}%
\fi\endgroup%
\begin{picture}(5523,2652)(451,-3743)
\put(871,-3001){\makebox(0,0)[lb]{\smash{\SetFigFont{7}{8.4}{\rmdefault}{\mddefault}{\updefault}$\Nthree(v)$}}}
\put(871,-2416){\makebox(0,0)[lb]{\smash{\SetFigFont{7}{8.4}{\rmdefault}{\mddefault}{\updefault}$\Ntwo(v)$}}}
\put(886,-1876){\makebox(0,0)[lb]{\smash{\SetFigFont{7}{8.4}{\rmdefault}{\mddefault}{\updefault}$\None(v)$}}}
\put(3331,-1996){\makebox(0,0)[lb]{\smash{\SetFigFont{9}{10.8}{\rmdefault}{\mddefault}{\updefault}$v$}}}
\end{picture}%
\quad
\begin{picture}(0,0)%
\epsfig{file=rule2.pstex}%
\end{picture}%
\setlength{\unitlength}{1450sp}%
\begingroup\makeatletter\ifx\SetFigFont\undefined%
\gdef\SetFigFont#1#2#3#4#5{%
  \reset@font\fontsize{#1}{#2pt}%
  \fontfamily{#3}\fontseries{#4}\fontshape{#5}%
  \selectfont}%
\fi\endgroup%
\begin{picture}(6930,3257)(541,-3199)
\put(7471,-2131){\makebox(0,0)[rb]{\smash{\SetFigFont{7}{8.4}{\rmdefault}{\mddefault}{\updefault}$\Nthree(v,w)$}}}
\put(7471,-1591){\makebox(0,0)[rb]{\smash{\SetFigFont{7}{8.4}{\rmdefault}{\mddefault}{\updefault}$\Ntwo(v,w)$}}}
\put(7471,-1006){\makebox(0,0)[rb]{\smash{\SetFigFont{7}{8.4}{\rmdefault}{\mddefault}{\updefault}$\None(v,w)$}}}
\put(541,-1816){\makebox(0,0)[lb]{\smash{\SetFigFont{9}{10.8}{\rmdefault}{\mddefault}{\updefault}$v$}}}
\put(4456,-1501){\makebox(0,0)[lb]{\smash{\SetFigFont{9}{10.8}{\rmdefault}{\mddefault}{\updefault}$w$}}}
\end{picture}%
\end{center}     
\caption{\label{fig:rules}
The left-hand side shows the partitioning of the neighborhood of 
a single vertex~$v$. The right-hand side shows the partitioning of
a neighborhood~$N(v,w)$ of two vertices~$v$ and~$w$. 
%
Since, in the left-hand figure,~$\Nthree(v)\not=\emptyset$, reduction Rule~\ref{rule1}
applies. In the right-hand figure, since~$\Nthree(v,w)$ cannot be dominated
by a single vertex at all, Case 2 of Rule~\ref{rule2} applies.}
\end{figure}

An example which illustrates the partitioning of~$N(v)$ into
the subsets~$\None(v)$, $\Ntwo(v)$, and $\Nthree(v)$ 
can be seen in the left-hand diagram of Fig.~\ref{fig:rules}.

Note that, by definition of the three subsets, the vertices in~$\Nthree(v)$ 
cannot be dominated by vertices from~$\None(v)$. A good candidate for 
dominating~$\Nthree(v)$ is given by the choice of~$v$.
Observing that this indeed is always an optimal choice lies the 
base for our first reduction rule.

\begin{rul}
\label{rule1}
If $\Nthree(v) \not= \emptyset$ for some vertex~$v$, then 
\begin{itemize}
\item remove $\Ntwo(v)$ and
$\Nthree(v)$ from~$G$ and 
\item add a new vertex $v'$ with the edge $\{v,v'\}$ to~$G$.
\end{itemize}
\end{rul}

We use the vertex~$v'$ as a ``gadget vertex'' that enforces us
to take $v$ (or $v'$) into an optimal dominating set in the reduced graph.

\begin{lemma}\label{rule1correct}
Let $G=(V,E)$ be a graph and let $G'=(V',E')$ be the 
resulting graph after having applied Rule~\ref{rule1}
to~$G$. Then $\gamma(G)=\gamma(G')$.
\end{lemma}
\begin{proof}
Consider a vertex~$v \in V$ such that $\Nthree(v) \not= \emptyset$.
The vertices in $\Nthree(v)$ can only be dominated
by either~$v$ or by vertices in $\Ntwo(v) \cup \Nthree(v)$. 
But, clearly, $N(w) \subseteq N(v)$ for every $w \in \Ntwo(v) \cup \Nthree(v)$.
This shows that an optimal way to dominate~$\Nthree(v)$ is given
by taking~$v$ into the dominating set. This is simulated~by the 
``gadget vertex'' $v'$ in $G'$ which enforces us 
to take $v$ (or $v'$) into an optimal dominating set. 
It is safe to remove~$\Ntwo(v)\cup \Nthree(v)$ since 
$N(\Ntwo(v)\cup \Nthree(v)) \subseteq N(v)$,
i.e., since the vertices that could be dominated by vertices from $\Ntwo(v)\cup \Nthree(v)$ are already
dominated by~$v$.
Hence, $\gamma(G') = \gamma(G)$. 
\end{proof}

\begin{lemma}
\label{time:rule1}
Rule~\ref{rule1} can be carried out in 
time~$O(n)$ for planar graphs and in time~$O(n^3)$
for general graphs. 
\end{lemma}
\begin{proof}
We first discuss the planar case.
To carry out Rule~1, for each vertex~$v$ of the given planar graph~$G$ 
we have to determine the neighbor sets $\None (v)$, $\Ntwo (v)$,
and $\Nthree (v)$.
By definition of these sets, one easily observes that it is sufficient 
to consider the subgraph~$G$ that is induced by all vertices 
that are connected to~$v$ by a path of length at most two.
To do so, we employ a search tree of depth two, rooted at~$v$.
We perform two phases.

In phase~1, constructing the search tree we determine the vertices 
from $\None (v)$.
Each vertex of the first level (i.e., distance one from the root~$v$)
of the search tree that has a neighbor at the second level
of the search tree belongs to $\None (v)$.
Observe that it is enough to stop the expansion of a vertex from the first 
level as soon as its {\em first\/} neighbor in the second level is 
encountered. Hence, denoting the degree of~$v$ by~$\deg(v)$, 
phase~1 takes time $O(\deg (v))$ because 
there clearly are at most $2\cdot\deg (v)$ tree edges 
and at most $O(\deg (v))$ non-tree edges to be explored.
The latter holds true since these non-tree edges all belong 
to the subgraph of~$G$ induced by~$N[v]$. Since this graph 
is clearly planar and $|N[v]| = \deg (v)+1$, the claim follows.

In phase 2, it remains to determine the sets $\Ntwo(v)$ and~$\Nthree(v)$.
To get $\Ntwo(v)$, one basically has to go through all vertices from the first
level of the above search tree that are not already marked as being 
in~$\None (v)$ but have at least one neighbor in~$\None (v)$.
All this can be done within the planar graph induced by~$N[v]$, using 
the already marked $\None (v)$-vertices, in time $O(\deg (v))$.
Finally, $\Nthree (v)$ simply consists of vertices from the first level that 
are neither marked being in~$\None (v)$ nor marked being in~$\Ntwo (v)$.
In summary, this shows that for vertex~$v$ the sets 
$\None (v)$, $\Ntwo (v)$, and $\Nthree (v)$ can be constructed in time
$O(\deg (v))$. 

Once having determined these three sets, the sizes of which all are bounded 
by~$\deg (v)$, it is clear that the possible removal of vertices 
from $\Ntwo (v)$ and~$\Nthree (v)$ and the addition of a vertex 
and an edge as required by Rule~1 all can be done 
in time $O(\deg (v))$.
Finally, it remains to analyze the overall complexity 
of this procedure when going through all $n$~vertices 
of~$G=(V,E)$.
But this is easy.
The running time can be bounded by 
$\sum_{v\in V} O(\deg (v))$.
Since $G$ is planar, this sum is bounded by~$O(n)$,
i.e., the whole reduction takes linear time.

For general graphs, 
the method described above leads to a worst-case cubic time 
implementation of Rule~1. Here, one ends up with the sum
$$\sum_{v\in V} O((\deg (v))^2) = O(n^3).$$ Note that the
size of the graph that is induced by the neighborhood~$N[v]$ 
again is relevant for the time needed to determine the sets~$\None(v)$, $\Ntwo(v)$, and $\Nthree(v)$. 
For general graphs, this neighborhood may contain~$O(\deg(v)^2)$ many vertices. 
\end{proof}

\subsection{The Neighborhood of a Pair of Vertices}
Similar to Rule~1, we explore
the neighborhood set $N(v,w):= N(v) \cup N(w)$ of two vertices $v,w \in V$.
Analogously, we now partition~$N(v,w)$ into
three disjoint subsets $\None(v,w)$, $\Ntwo(v,w)$, and $\Nthree(v,w)$.
Setting $N[v,w]:= N[v]\cup N[w]$, we define
\begin{eqnarray*}
\None(v,w) & := & \{ u \in N(v,w) : N(u) \setminus N[v,w] \not= \emptyset \} , \\
\Ntwo(v,w) & := & \{ u \in N(v,w)\setminus \None(v,w) \mid N(u) : \None(v,w) \not= \emptyset \} ,\\
\Nthree(v,w) & := & N(v,w) \setminus (\None(v,w) \cup \Ntwo(v,w)).
\end{eqnarray*}

The right-hand diagram of Fig.~\ref{fig:rules} shows 
an example which illustrates the partitioning of~$N(v,w)$ into
the subsets~$\None(v,w)$, $\Ntwo(v,w)$, and $\Nthree(v,w)$.

Our second reduction rule---compared to Rule~\ref{rule1}---is
slightly more complicated.

\begin{rul}
\label{rule2}
Consider 
$v,w \in V$ ($v \not= w$) and suppose that $\Nthree(v,w)\not= \emptyset$.
Suppose that $\Nthree(v,w)$ cannot be dominated
by a single vertex from $\Ntwo(v,w) \cup \Nthree(v,w)$.
\begin{description}
\item[Case 1]
If $\Nthree(v,w)$ can be dominated by a single vertex from $\{v,w\}$:
  \begin{description}
     \item[(1.1)] If $\Nthree(v,w) \subseteq N(v)$ as well as $\Nthree(v,w) \subseteq N(w)$: 
    \begin{itemize}
      \item remove $\Nthree(v,w)$ and $\Ntwo(v,w)\cap N(v) \cap N(w)$ from~$G$ and
      \item add two new vertices $z,z'$ and edges $\{v,z\}$, $\{w,z\}$, $\{v,z'\}$, $\{w,z'\}$ to~$G$. 
    \end{itemize}
     \item[(1.2)] If $\Nthree(v,w) \subseteq N(v)$, but not $\Nthree(v,w) \subseteq N(w)$:    
       \begin{itemize}
       \item remove $\Nthree(v,w)$ and $\Ntwo(v,w) \cap N(v)$ from~$G$ and
       \item add a new vertex $v'$ and the edge $\{v,v'\}$ to~$G$. 
       \end{itemize}
     \item[(1.3)] If $\Nthree(v,w) \subseteq N(w)$, but not $\Nthree(v,w) \subseteq N(v)$:
       \begin{itemize}
         \item remove $\Nthree(v,w)$ and $\Ntwo(v,w) \cap N(w)$ from~$G$ and
         \item add a new vertex $w'$ and the edge $\{w,w'\}$ to~$G$. 
       \end{itemize}
  \end{description}
\item[Case 2]
If $\Nthree(v,w)$ cannot be dominated by a single vertex from $\{v,w\}$:
  \begin{itemize}
     \item remove $\Nthree(v,w)$ and $\Ntwo(v,w)$ from~$G$ and
     \item add two new vertices $v',w'$ and edges $\{v,v'\}$, $\{w,w'\}$ to~$G$.
  \end{itemize}
 \end{description}
\end{rul}

Again, the newly added vertices $v'$ and~$w'$ of degree one act as gadgets that enforce
us to take $v$ or $w$ into an optimal dominating set. A special situation is given in
Case~(1.1). Here, the gadet added to the graph~$G$ simulates that at least one of the vertices
$v$ or $w$ has to be taken into an optimal dominating set. 

\begin{lemma}
\label{lem:correct2}
Let $G=(V,E)$ be a graph and let $G'=(V',E')$ be the 
resulting graph after having applied Rule~\ref{rule2}
to~$G$. Then $\gamma(G)=\gamma(G')$. 
\end{lemma}
\begin{proof}
Similar to the proof of Lemma~\ref{rule1correct},
we observe that vertices from~$\Nthree(v,w)$ can only be dominated
by vertices from~$M:=\{v,w\} \cup \Ntwo(v,w) \cup \Nthree(v,w)$.
All cases in Rule~\ref{rule2}
are based on the fact that~$\Nthree(v,w)$ needs to be dominated.
All cases only apply if there is not a {\em single} vertex in
$\Ntwo(v,w) \cup \Nthree(v,w)$ which dominates~$\Nthree(v,w)$.

We first of all discuss the correctness of Case~(1.2) (and
similarly obtain the correctness of the symmetric Case~(1.3)): 
If $v$ dominates 
$\Nthree(v,w)$ (and~$w$ does not) then it is optimal to take~$v$ 
into the dominating set---and at the same time still leave the option of 
taking vertex~$w$---than to take any combination of two vertices~$x,y$
from the set~$M\setminus\{v\}$. It may be that we still have to take~$w$
to get a minimum dominating set, but in any case $v$ and $w$ dominate at least
as many vertices as~$x$ and~$y$.
 The ``gadget edge'' $\{v,v'\}$ simulates the effect of taking~$v$.
It is safe to remove~$R:=(\Ntwo(v,w)\cap N(v)) \cup \Nthree(v,w)$ since, by taking 
$v$ into the dominating set, all vertices in~$R$ are already dominated and since, as discussed above,
it is always at least as good to take~$\{v,w\}$ into a minimum  dominating set
than to take $v$ and any other of the vertices from~$R$.

In the situation of Case~(1.1), we can dominate~$\Nthree(v,w)$ by
both either~$v$ or~$w$. Since we cannot decide at this point which of these
vertices should be chosen to be in the dominating set, we use the
gadget with vertices~$z$ and~$z'$ which simulates a choice
between $v$ or~$w$, as can be seen easily.
In any case, however, it is at least as good to take one
of the vertices $v$ and~$w$ (maybe both) than to take any two 
vertices from~$M \setminus\{v,w\}$. The argument for this is similar
to the one for Case~(1.2). The removal of 
$\Nthree(v,w) \cup (\Ntwo(v,w)\cap N(v) \cap N(w))$ is safe by a similar
argument as the one that justified the removal of~$R$ in Case~(1.2).

Finally, in Case 2, we clearly need at least two vertices
to dominate~$\Nthree(v,w)$. Since $N(v,w) \supseteq N(x,y)$ for 
all pairs $x,y \in M$ it is optimal to take $v$ and~$w$ into the dominating
set, simulated by the gadgets~$\{v,v'\}$ and~$\{w,w'\}$.
As in the previous cases the removal of~$\Nthree(v,w) \cup \Ntwo(v,w)$ is
safe since these vertices are already dominated and since these vertices
need not be used for an optimal dominating~set.
\end{proof}

\begin{lemma}
\label{time:rule2}
Rule~\ref{rule2} can be carried out in 
time~$O(n^2)$ for planar graphs and in 
time~$O(n^4)$ for general graphs. 
\end{lemma}
\begin{proof}
To prove the time bounds for Rule~2, basically the same ideas as for Rule~1 apply (cf.\ proof of 
Lemma~\ref{time:rule1}). Instead of a depth two search tree,
one now has to argue on a search tree where the levels indicate the minimum of the distances
to vertex $v$ and~$w$. Hence, we associate the vertices~$v$ and~$w$ to the root of this search tree.
The first level consists of all vertices that lie in~$N(v,w)$ (i.e., at distance
one from either of the vertices~$v$ or~$w$). Determining the subset~$\Nthree(v,w)$ 
means to check whether some vertex on the first level has a neighbor on the second level.
We do the same kind of construction as in Lemma~\ref{time:rule1}. The running time again
is determined by the size of the subgraph induces by the vertices that correspond to the root and
the first level of this search tree, i.e., by~$G[N[v,w]]$ in this case. 
For planar graphs, we have $|G[N[v,w]]| = O\left(\deg (v)+\deg (w)\right)$.
Hence, we get 
$\sum_{v,w\in V} O\left(\deg (v)+\deg (w)\right)$ as an upper bound 
on the overall running time in the case of planar graphs.
This is upperbounded by 
$$O(\sum_{v\in V} (n\cdot\deg (v) + \sum_{w\in V} \deg (w))) = O(n^2).$$
In case of general graphs, we have $|G[N[v,w]]| = O\left((\deg (v)+\deg (w))^2\right)$, 
which trivially yields the upper bound
$$\sum_{v,w\in V} O((\deg (v) +\deg (w))^2) = O(n^4)$$
for the overall running time. 
\end{proof}

We remark that the running times given in Lemmas~\ref{time:rule1} and~\ref{time:rule2}
are pure worst-case estimates and turn out to be much lower in our experimental studies.
In particular, for practical purposes it is important to see 
that Rule~\ref{rule2} can only be applied for vertex pairs 
that are at distance at most three. The algorithms implementing these rules appear
to be much faster (see the Section~\ref{sec:concl}).

\subsection{Reduced Graphs}

\begin{definition}
Let $G=(V,E)$ be a graph such that 
both the application of Rule~1 and the application
of Rule~2 leave the graph unchanged.
Then we say that $G$ is
{\em reduced} with respect to these rules.
\end{definition}

Observing that the (successful) application of any reduction rule
always ``shrinks'' the given graph implies that there can
only be $O(n)$ successful applications of reduction rules.
This leads to the following.

\begin{theorem}
\label{time:total}
A graph $G$ can be transformed into 
a reduced graph $G'$ with $\gamma(G)=\gamma(G')$
in time~$O(n^3)$ in the planar case and in time $O(n^5)$ 
in the general case.
\end{theorem}

\begin{remark}
{\rm The algorithms presented in the proofs of Lemmas are very simple to implement
and behave well in practice.
{F}rom a theoretical point of view, using the concept of tree decompositions
it might even be possible to transform a planar graph~$G$ into a reduced graph
asymptotically faster\cite{Bod02}. The basic observation is that the application of a rule only changes
the graph locally (i.e., it affects vertices that are at most at distance five 
from each other). Due to the involved constant factors of such a tree decomposition
based algorithm, however, this approach is impractical.}
%
%
\end{remark}

%
\begin{remark}
\label{rem:rule1and2}
{\rm A graph~$G=(V,E)$ which is reduced with respect to reduction 
Rules~\ref{rule1} and~\ref{rule2} has the following properties:
\begin{enumerate}
\item For all $v \in V$, the set~$\Nthree(v)$ is always empty (these 
vertices are removed by Rule~\ref{rule1}) except for it may contain a
single gadget vertex of degree one. 
\item For all $v,w \in V$, there exists a single vertex in $\Ntwo(v,w) \cup \Nthree(v,w)$
which dominates all vertices~$\Nthree(v,w)$
(in all other cases Rule~\ref{rule2} is applied).
\end{enumerate}}
\end{remark}

\section{A Linear Problem Kernel for Planar Graphs}
Here, we show that the reduction rules given in
Section~\ref{rules} yield a linear size
problem kernel for {\sc dominating set} 
on planar graphs. 
Such a result is very unlikely to hold for general graphs,
since {\sc dominating set} is W[2]-complete and the
existence of a (linear) problem 
kernel implies fixed-parameter tractability.


\begin{theorem}
\label{thm}
For a planar graph $G=(V,E)$ which is reduced 
with respect to Rules~\ref{rule1} and~\ref{rule2}, we
get $|V| \le 335 \, \gamma(G)$, 
i.e., the {\sc dominating set} problem on planar graphs 
admits a linear problem kernel.
\end{theorem}

The rest of this section is devoted to the proof of 
Theorem~\ref{thm}. 
The proof can be split into two parts.
In a first step, we try to find a so-called ``maximal region decomposition''
of the vertices~$V$ of a reduced graph~$G$. 
In a second step, we show, on the one hand, 
that such a maximal region decomposition
must contain all but~$O(\gamma(G))$ many 
vertices from~$V$.
On the other hand, we prove that
such a region decomposition uses at 
most~$O(\gamma(G))$ regions, each of which having size~$O(1)$.
Combining the results then yields $|V| = O(\gamma(G))$.

The notion of ``region decompositions'' heavily relies on 
the planarity of our input graph and cannot be carried over
to general graphs.

\subsection{Finding a Maximal Region Decomposition}

Suppose that we have a reduced planar graph~$G$ with a minimum
dominating set~$D$. We know that, in particular, 
neither Rule~\ref{rule1} applies to 
a vertex~$v\in D$ nor Rule~\ref{rule2} applies to 
a pair of vertices $v,w \in D$. 
We want to get our hands on the number of vertices
which lie in neighborhoods $N(v)$ for~$v \in D$, or neighborhoods~$N(v,w)$ 
for $v,w \in D$. A first idea to prove that $|V| = O(|D|)$
would be to find ($\ell=O(|D|)$ many) neighborhoods 
$N(v_1,w_1), \ldots, N(v_\ell,w_\ell)$ with $v_i,w_i \in D$ such 
that all vertices in~$V$ lie in at least one such neighborhood;
and then use the fact that~$G$ is reduced in order to
prove that each~$N(v_i,w_i)$ has size~$O(1)$.
Even if the graph~$G$ is reduced, however, the neighborhoods~$N(v,w)$
of two vertices~$v,w \in D$ may contain many vertices: the size of~$N(v,w)$ in a reduced
graph basically
depends on how big~$\None(v,w)$ is. 

In order to circumvent these difficulties, we define 
the concept of a region~$R(v,w)$ for which we can 
guarantee that in a reduced graph it consists of only a
constant number of vertices.



\begin{definition}
\label{def:1-reg}
Let $G=(V,E)$ be a plane\footnote{A plane graph is a particular 
planar embedding of a planar graph.} graph.
A {\em region}
$R(v,w)$ between two vertices $v,w$ is 
a closed subset of the plane with the following properties:
\begin{enumerate}
\item the boundary of $R(v,w)$ is formed by two simple paths~$P_1$ 
and~$P_2$ in~$V$ which connect $v$ and~$w$, and the length of
each path is at most three\footnote{The length of a path is the number of edges on it.}, and
\item all vertices which are strictly inside\footnote{%
By ``strictly inside the region~$R(v,w)$'' we mean lying in the region, but not sitting on the boundary of $R(v,w)$.} 
the region~$R(v,w)$ are from~$N(v,w)$. 
\end{enumerate}

For a region~$R=R(v,w)$, let $V(R)$ denote the vertices 
belonging to~$R$, i.e., 
$$
V(R) := \{ u \in V \mid \text{$u$ sits inside or on the boundary of~$R$} \}.
$$
\end{definition}

\begin{definition}
\label{def:sep}
Let~$G=(V,E)$ be a plane graph and~$D \subseteq V$. 
A {\em $D$-region decomposition} of~$G$ is a set~$\cal R$ 
of regions between pairs of vertices in~$D$ such that 
\begin{enumerate}
\item 
for~$R(v,w) \in {\cal R}$ 
no vertex from~$D$ (except for~$v,w$) lies in~$V(R(v,w))$ and
\item 
no two regions $R_1,R_2  \in {\cal R}$ do intersect 
(however, they may touch each other by having common boundaries).
\end{enumerate}
For a $D$-region decomposition~$\cal R$, we define
$V({\cal R}):=\bigcup_{R \in \cal R} V(R)$.
 A $D$-region decomposition~$\cal R$ is called {\em maximal} if 
there is no region~$R \notin \cal R$ such that
${\cal R'} := {\cal R} \cup \{R\}$ is a $D$-region decomposition with 
$V({\cal R}) \subsetneq V({\cal R'})$.
\end{definition}

\begin{figure}[t]
\begin{center}
\begin{minipage}{5.1cm}
\mbox{}\epsfxsize5cm\epsfbox{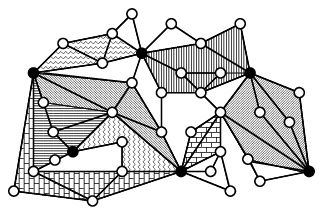} 
\end{minipage}
\quad
\begin{minipage}{3.1cm}
\mbox{}\epsfxsize3cm\epsfbox{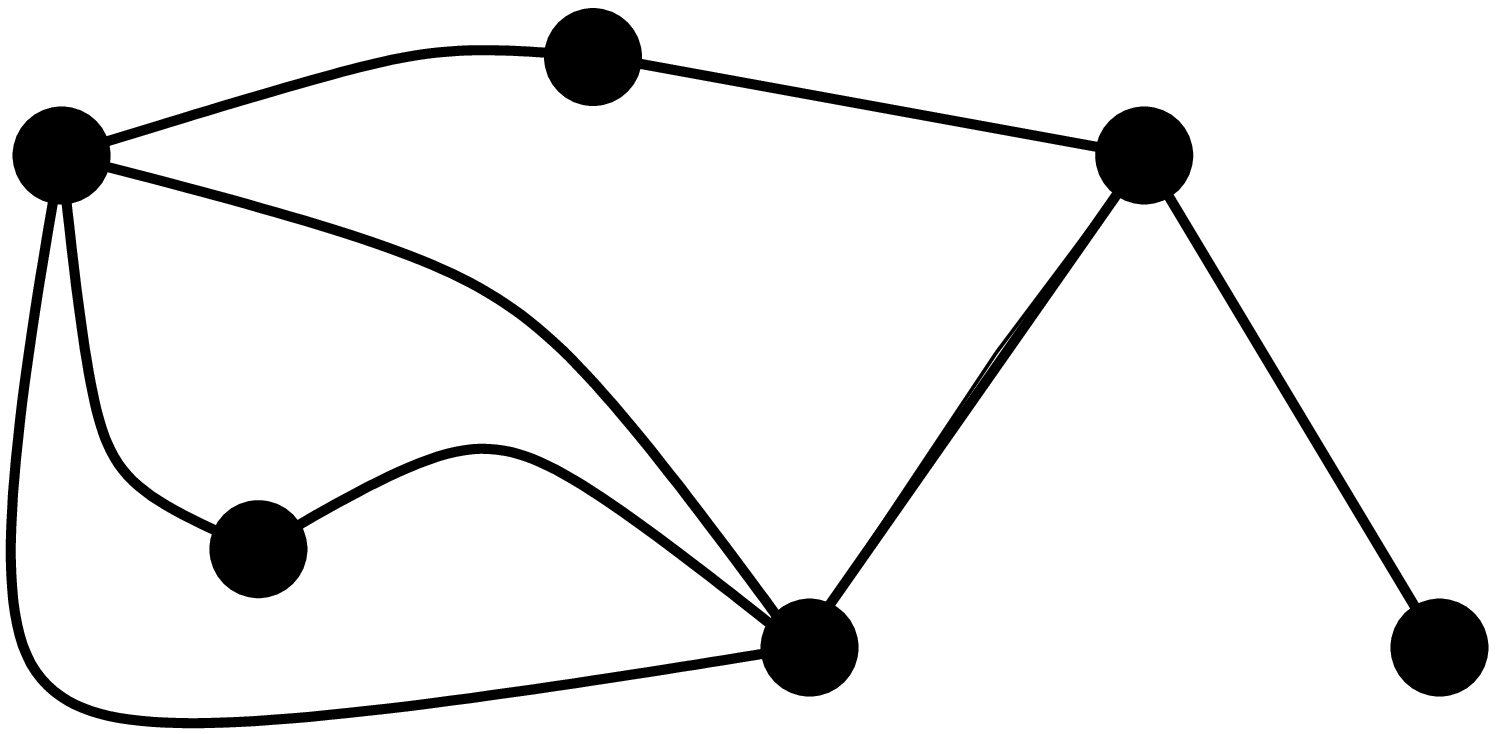} 
\end{minipage}
\end{center}     
\caption{\label{fig:dec}
The left-hand side diagram shows an example of a possible $D$-region 
decomposition~$\cal R$ of some graph~$G$, where
$D$ is the subset of vertices in~$G$ that
are drawn in black. The various regions are highlightened
by different patterns.
The remaining white areas are not considered as regions.
The given $D$-region decomposition 
is maximal.
The right-hand side shows the induced graph~$G_{{\cal R}}$ (Definition~\ref{def:indgraph}).}
\end{figure}

For an example of a (maximal) $D$-region decomposition we refer to the 
left-hand side diagram of Fig.~\ref{fig:dec}.


We will show that, for a given graph~$G$ with 
dominating set~$D$, we can always find a maximal $D$-region decomposition
with at most~$O(\gamma(G))$ many regions.
For that purpose, we observe that a $D$-region decomposition induces 
a graph in a very natural way.

\begin{definition}
\label{def:indgraph}
The {\em induced graph}~$G_{\cal R}=(V_{\cal R},E_{\cal R})$ of a $D$-region 
decomposition~$\cal R$ of~$G$ is the graph with possible multiple edges which 
is defined by 
$V_{\cal R} :=  D$  and
\begin{equation*}
 E_{\cal R} :=  \{ \{v,w\} \mid 
\text{there is a region~$R(v,w) \in \cal R$ between $v,w \in D$} \}.
\end{equation*}
\end{definition}

Note that, by~Definition~\ref{def:sep},
the induced graph~$G_{\cal R}$ of a $D$-region decomposition is planar. For an 
example of an induced graph~$G_{\cal R}$ see Fig.~\ref{fig:dec}.

\begin{definition}
A planar graph~$G=(V,E)$ with multiple edges 
is {\em thin}
if there exists a planar embedding such that no two multiedges are homotopic: This means that 
if there are two edges $e_1, e_2$ between a pair of 
distinct vertices $v,w \in V$, then there must be two further vertices 
$u_1,u_2 \in V$ which sit inside the two disjoint areas of the plane
that are enclosed by $e_1,e_2$.
\end{definition}

The induced graph~$G_{\cal R}$ in Fig.~\ref{fig:dec} is thin.

\begin{lemma}
\label{lem:thin}
For a thin planar graph $G=(V,E)$ we have
$|E|  \le  3|V| -6$.
\end{lemma}
\begin{proof}
The claim is true for planar graphs without multiple edges.
An easy induction on the number of multiple edges in~$G$ proves the claim.
\end{proof}

Using the notion of thin graphs, we can formulate the main result 
of this subsection.
\begin{proposition}
\label{prop:dec}
For a reduced plane graph~$G$ with dominating set~$D$, 
there exists a maximal $D$-region decomposition~$\cal R$
such that $G_{\cal R}$ is thin. 
\end{proposition}
%
\begin{figure}[t]
\mbox{\begin{minipage}{\textwidth}
\small
\noindent{\tt region\_decomp(plane graph $G=(V,E)$, vertex subset $D \subseteq V$)}\\
// Returns a $D$-region decomposition~$\cal R$ for $G$ such that \\
// the induced graph $G_{\cal R}$ is thin.
\begin{itemize}
\item Let $V_{\text{used}} \leftarrow \emptyset$; $\mathcal{R} \leftarrow \emptyset$.
\item {\tt For all $u \in V$ do}
  \begin{itemize}
  \item {\tt If} (($u \notin V_{\text{used}}$) {\tt and} ($u \in V(R)$ for some region
        $R=R(v,w)$~between\\ two vertices $v,w \in D$ such that ${\cal R} \cup \{R\}$ is a $D$-region decomposition)) {\tt then}
      \begin{itemize}
      \item Consider the set~${\cal R}_u$ of all regions~$S$ with the following 
properties:\footnote{%
These four properties ensure that ${\cal R} \cup \{S\}$ is a 
$D$-region decomposition for every $S \in {\cal R}_u$.}
            \begin{enumerate}
            \item $S$ is a region between $v$ and $w$.
            \item $S$ contains $u$.
            \item no vertex from $D\setminus \{v,w\}$ is in~$V(S)$.
            \item $S$ does not cross any region from~$\cal R$. 
            \end{enumerate}
           \item Choose a region $S_u \in {\cal R}_u$ which is maximal in space.\footnote{%
A region~$S_u$ is maximal in space if $S'\supseteq S_u$ for any $S' \in {\cal R}_u$ implies $S'=S_u$.}
            \item ${\cal R} \leftarrow {\cal R} \cup \{S_u\}$.
            \item $V_{\text{used}} \leftarrow V_{\text{used}} \cup V(S_u)$.
        \end{itemize} 
    \end{itemize} 
\item {\tt Return~$\cal R$.}
\end{itemize}
\end{minipage}}
\caption{\label{fig:regdec} Greedy-like construction of a maximal $D$-region decomposition.}
\end{figure}
%
\begin{proof}
We give a constructive proof on how to find a maximal $D$-region 
decomposition~$\cal R$ of a plane graph~$G$ such that the induced graph~$G_{\cal R}$
is thin. Consider the algorithm presented in Fig.~\ref{fig:regdec}.
It is obvious that the algorithm returns a $D$-region decomposition,
since---by construction---we made sure that regions are between vertices
in~$D$, that regions do not contain vertices from~$D$, and that regions
do not intersect. Moreover, the $D$-region decomposition obtained by the 
algorithm is maximal: If a vertex~$u$ does not belong to 
a region, i.e., if $u \notin V_{\text{used}}$, then the algorithm eventually 
checks,
whether there is a region~$S_u$ such that ${\cal R} \cup \{S_u\}$ is 
a $D$-region decomposition. 

It remains to show that the induced graph~$G_{\cal R}$ of the 
$D$-region decomposition~$\cal R$ found by the algorithm is thin. 
We embed~$G_{\cal R}$ in the plane in such a way that an edge belonging to 
a region~$R \in \cal R$ is drawn inside the area covered by~$R$. 
To see that the graph is thin, we have to show that, 
for every multiple edge~$e_1,e_2$ (belonging to two regions~$R_1,R_2 \in \cal R$
that were chosen at some point of the algorithm)
between two vertices $v,w\in D$, there exist two vertices~$u_1,u_2 \in D$
which lie inside the areas enclosed by~$e_1,e_2$. Let~$A$ be such an area.
Suppose that there is no vertex~$u \in D$ in~$A$.
We distinguish two cases. Either there is also no vertex from~$V\setminus D$
in~$A$ or there are other vertices~$V'$ from~$V\setminus D$ inside~$A$. 
In the first case, by joining the regions~$R_1$ and~$R_2$ we obtain
a bigger region which fulfills all the four conditions checked
by the algorithm in Fig.~\ref{fig:regdec}, a contradiction to 
the maximality of $R_1$ and $R_2$.
In the second case, since~$D$ is assumed to be a dominating set,
the vertices in~$V'$ need to be dominated by~$D$. Since $v,w$ are the 
only vertices from~$D$ which are part of~$A$, $R_1$ or $R_2$, the
vertices in~$V'$ need to be dominated by $v,w$, hence they belong 
to~$N(v,w)$. 
But then again by joining the regions~$R_1$ and~$R_2$ we obtain
a bigger region which again fulfills all the four conditions 
of the algorithm in Fig.~\ref{fig:regdec}, a contradiction to 
the maximality of $R_1$ and $R_2$.
\end{proof}

\subsection{Region Decompositions and the Size of Reduced Planar Graphs}

Suppose that we are given a reduced planar graph~$G=(V,E)$ 
with a minimum dominating
set~$D$. 
Then, by Proposition~\ref{prop:dec} and Lemma~\ref{lem:thin}, 
we can find a maximal $D$-region 
decomposition~$\cal R$ of~$G$ with at most $O(\gamma(G))$ regions. 
In order to see that $|V| = O(\gamma(G))$, it remains to show that 
\begin{enumerate}
\item 
there are at most $O(\gamma(G))$ vertices which do not belong
to any of the regions in~$\cal R$, and that
\item 
every region of~$\cal R$ contains at most~$O(1)$ vertices.
\end{enumerate}
These issues are treated by the following two propositions.

We first of all 
state two technical lemmas, one which characterizes an important property of a maximal 
region decomposition and another one which gives an upper bound on 
the size of a special type of a region.

\begin{lemma}
\label{lem:maxproperty}
Let~$G$ be a reduced planar graph with a dominating set~$D$  
and let~$\cal R$ be a maximal $D$-region decomposition.
If $u \in \None(v)$ for some vertex~$v \in D$ then
$u \in V({\cal R})$.
\end{lemma}
\begin{proof}
Let $u\in \None(v)$ for some~$v \in D$ and assume that 
$u \notin V({\cal R})$. By definition of~$\None(v)$,
there exists a vertex $u'\in N(u)$ with $u' \notin N[v]$.
We distinguish two cases. Either~$u' \in D$ or 
$u'$ needs to be dominated by a vertex~$w\in D$ with~$w\not=v$.
If $u' \in D$, we consider the (degenerated) region consisting of the
path~$\{v,u,u'\}$. Since ${\cal R}$ is assumed to be maximal, this path
must cross a region~$R \in {\cal R}$. But this implies that~$u \in V(R)$, 
a contradiction.

In the second case, 
we consider the (degenerated) region consisting
of the path $\langle v,u,u',w \rangle$. Again, 
by maximality of~${\cal R}$, this path 
must cross a region~$R=R(x,y) \in {\cal R}$ between 
two vertices~$x,y \in D$.
Since, by assumption, $u \notin V(R)$, the edge~$\{u',w\}$ has to cross~$R$
which implies that~$w$ lies on the boundary of or inside~$R$ and, hence, $w\in V(R)$. 
However, according to the definition of a $D$-region decomposition, the only 
vertices from~$D$ that are in~$V(R)$ are~$x,y$. Hence, w.l.o.g., $x=w$.
At the same time~$u'$ must lie on the boundary of~$R$, otherwise~$u \in V(R)$.
By definition of a region, there exists path~$P$ of length 
at most three between~$w$ and~$y$ that goes through~$u'$ and that is part of the boundary of~$R$.
We claim that~$u'$ is a neighbor of~$y$: If this were not the case, 
the edge~$\{u',w~\}$ would be {\em on}~$P$. We already remarked, however, that the
edge~$\{u',w~\}$ {\em crosses}~$R$ and, thus, cannot lie on the boundary,
a contradiction to $u'$ not being neighbor of~$y$.
We know that~$u' \notin N(v)$, hence, $y\not=v$.
But then, 
the (degenerated) region~$R'$ consisting of the 
path~$\{v,u,u',y\}$ is a region between two vertices $v$ and $y$ in~$D$,
which does not cross (it only touches~$R$) any region in~${\cal R}$.
For the $D$-region decomposition ${\cal R}':={\cal R}\cup \{R'\}$, we have
$u \in V({\cal R'}) \setminus V({\cal R})$, contradicting the 
maximality of~$\cal R$.
\end{proof}

We now investigate a special type of a region
specified by the following definition. 

\begin{definition}
A region $R(v,w)$ between two vertices~$v,w \in D$ is called {\em simple} 
if all vertices contained in~$R(v,w)$ except for $v,w$ are common neighbors 
of both~$v$ and~$w$, 
i.e., if
$(V(R(v,w))\setminus\{v,w\}) \subseteq N(v) \cap N(w)$.

Let $v,u_1,w,u_2$ be the vertices that sit on the boundary
of the simple region~$R(v,w)$. We say that $R(v,w)$ is a
simple region of {\em Type~$i$} $(0 \le i \le 2)$ if 
$i$~vertices from~$\{u_1,u_2\}$ have a neighbor outside~$R(v,w)$.
\end{definition}

\begin{figure}[t]
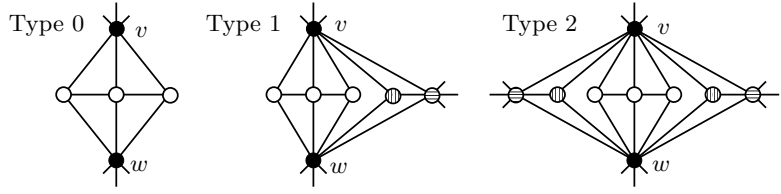

\begin{center}
\begin{picture}(0,0)%
\epsfig{file=simple_region1.pstex}%
\end{picture}%
\setlength{\unitlength}{1450sp}%
\begingroup\makeatletter\ifx\SetFigFont\undefined%
\gdef\SetFigFont#1#2#3#4#5{%
  \reset@font\fontsize{#1}{#2pt}%
  \fontfamily{#3}\fontseries{#4}\fontshape{#5}%
  \selectfont}%
\fi\endgroup%
\begin{picture}(2874,3216)(1126,-2794)
\put(3286,-151){\makebox(0,0)[lb]{\smash{\SetFigFont{9}{10.8}{\rmdefault}{\mddefault}{\updefault}$v$}}}
\put(3196,-2446){\makebox(0,0)[lb]{\smash{\SetFigFont{9}{10.8}{\rmdefault}{\mddefault}{\updefault}$w$}}}
\put(1126,-61){\makebox(0,0)[lb]{\smash{\SetFigFont{9}{10.8}{\rmdefault}{\mddefault}{\updefault}Type 0}}}
\end{picture}%
\quad 
\begin{picture}(0,0)%
\epsfig{file=simple_region2.pstex}%
\end{picture}%
\setlength{\unitlength}{1450sp}%
\begingroup\makeatletter\ifx\SetFigFont\undefined%
\gdef\SetFigFont#1#2#3#4#5{%
  \reset@font\fontsize{#1}{#2pt}%
  \fontfamily{#3}\fontseries{#4}\fontshape{#5}%
  \selectfont}%
\fi\endgroup%
\begin{picture}(4308,3216)(1126,-2794)
\put(3241,-2491){\makebox(0,0)[lb]{\smash{\SetFigFont{9}{10.8}{\rmdefault}{\mddefault}{\updefault}$w$}}}
\put(3331,-61){\makebox(0,0)[lb]{\smash{\SetFigFont{9}{10.8}{\rmdefault}{\mddefault}{\updefault}$v$}}}
\put(1126,-61){\makebox(0,0)[lb]{\smash{\SetFigFont{9}{10.8}{\rmdefault}{\mddefault}{\updefault}Type 1}}}
\end{picture}%
\quad
\begin{picture}(0,0)%
\epsfig{file=simple_region3.pstex}%
\end{picture}%
\setlength{\unitlength}{1450sp}%
\begingroup\makeatletter\ifx\SetFigFont\undefined%
\gdef\SetFigFont#1#2#3#4#5{%
  \reset@font\fontsize{#1}{#2pt}%
  \fontfamily{#3}\fontseries{#4}\fontshape{#5}%
  \selectfont}%
\fi\endgroup%
\begin{picture}(5016,3216)(418,-2794)
\put(3331,-106){\makebox(0,0)[lb]{\smash{\SetFigFont{9}{10.8}{\rmdefault}{\mddefault}{\updefault}$v$}}}
\put(3241,-2446){\makebox(0,0)[lb]{\smash{\SetFigFont{9}{10.8}{\rmdefault}{\mddefault}{\updefault}$w$}}}
\put(676,-61){\makebox(0,0)[lb]{\smash{\SetFigFont{9}{10.8}{\rmdefault}{\mddefault}{\updefault}Type 2}}}
\end{picture}%

 \caption{\label{fig:simplereg} Simple regions of Type 0, Type 1, Type 2.
This figure illustrates the largest possible simple regions in a reduced
graph. Vertices marked with horizontal lines are in~$\None(v,w)$, 
vertices marked with vertical lines belong to~$\Ntwo(v,w)$, and white
vertices are in~$\Nthree(v,w)$.}
\end{center}
\end{figure}

\begin{lemma}
\label{lem:simpleregion}
Every simple region~$R$ of Type~$i$ of a plane reduced 
graph contains at most $5+2i$ vertices.
\end{lemma}
\begin{proof}
Let~$R=R(v,w)$ be a simple region of Type~$i$ between vertices~$v$ and~$w$.
We will show that~$|V(R)| \le 5+2i$. The worst-case simple regions are 
depicted in Fig.~\ref{fig:simplereg}.
Firstly, let us count the number of vertices in~$V(R)$ which
belong to~$\None(v,w) \cup \Ntwo(v,w)$. Clearly, only vertices on the boundary
(except for~$v$ and~$w$) can have a neighbor outside~$R$. Thus,
all vertices in~$\None(v,w) \cap V(R)$ lie on the boundary of~$R$.
By definition of a simple 
region of Type~$i$, we have $|\None(v,w) \cap V(R)| \le i$.
Moreover, it is easy to see that, by planarity, every 
vertex in~$\None(v,w) \cap V(R)$ can contribute
at most one vertex to $\Ntwo(v,w) \cap V(R)$.
Hence, we get $|(\None(v,w)\cup \Ntwo(v,w)) \cap V(R)| \le 2i$

Secondly, we determine the number of vertices 
in~$\Nthree(v,w) \cap V(R)$.
Since~$G$ is reduced, by Remark~\ref{rem:rule1and2}, we know
that these vertices need to be dominated by a single
vertex in~$\Ntwo(v,w) \cup \Nthree(v,w)$. 
Moreover, since the region is simple,
all vertices in $\Nthree(v,w) \cap V(R)$ are neighbors of both $v$
and~$w$. By planarity, it follows that there can be at most
$3$~vertices in~$\Nthree(v,w) \cap V(R)$.

In summary, together with the vertices~$v,w \in V(R)$, 
we get $|V(R)| \le 5+2i$.
\end{proof}

We use Lemmas~\ref{lem:maxproperty} and~\ref{lem:simpleregion} for the following two proofs.

\begin{proposition}
\label{prop:leftovers}
Let~$G=(V,E)$ be a plane reduced graph and let $D$ be 
a dominating set of~$G$. If~$\cal R$ is a maximal
$D$-region decomposition then 
$
\left| V \setminus V({\cal R}) \right| \le 2|D|+ 56|{\cal R}|.
$
\end{proposition}
\begin{proof}
We claim that every vertex~$u \in V\setminus V({\cal R})$ is either a vertex 
in~$D$ or belongs to a set~$\Ntwo(v) \cup \Nthree(v)$ for some~$v \in D$. 
To see this, suppose that~$u \notin D$. But since~$D$ is a dominating
set, we know that $u \in N(v)=\None(v) \cup \Ntwo(v) \cup \Nthree(v)$
for some vertex $v \in D$.
Since~$\cal R$ is assumed to be maximal,
by Lemma~\ref{lem:maxproperty}, we know that $\None(v) \subseteq V({\cal R})$. 
Thus, $u \in \Ntwo(v) \cup \Nthree(v)$.  

For a vertex $v \in D$, let $\Ntwo^\ast(v)= \Ntwo(v) \setminus V({\cal R})$.
The above observation implies that~$V\setminus V({\cal R}) \subseteq D \cup (\bigcup_{v \in D}\Nthree(v)) \cup  (\bigcup_{v \in D} \Ntwo^\ast(v))$.

We, firstly, upperbound the size of $\bigcup_{v \in D} \Nthree(v)$.
Since, by Remark~\ref{rem:rule1and2},
$|\Nthree(v)| \le 1$, we get $|\bigcup_{v \in D} \Nthree(v) \le |D|$.

We now upperbound the size of~$\Ntwo^\ast(v)$ for a given vertex~$v \in D$.
To this end, for a vertex~$v \in D$, let $\None^\ast(v)$ be the subset of~$\None(v)$ 
which sit on the boundary of a region in~$\cal R$.
It is clear that $\Ntwo^\ast(v) \subseteq N(v) \cap N(\None^\ast(v))$. Hence, we 
investigate the set~$\None^\ast(v)$. Suppose that $R(v,w_1),\ldots,R(v,w_\ell)$ are the 
regions between~$v$ and some other vertices~$w_i \in D$, where $\ell=\text{deg}_{G_{\cal R}}(v)$ 
is the degree of~$v$ in the induced region graph~$G_{{\cal R}}$. 
Then, every region~$R(v,w_i)$ can contribute at most two vertices $u_i^1,u_i^2$ to $\None^\ast(v)$,
i.e., in the worst-case, we have $\None^\ast(v) = \bigcup_{i=1}^\ell \{u_i^1,u_i^2\}$ with
$u_i^1,u_i^2 \in V(R(v,w_i))$, i.e., $|\None^\ast(v)| \le 2 \deg_{G_{\cal R}}(v)$. 
%
We already observed that every vertex in~$\Ntwo^\ast(v)$ must be a 
common neighbor of~$v$ and some vertex in $\None^\ast(v)$.
We claim that, moreover, the vertices
in~$\Ntwo^\ast(v)$ can be grouped into various simple regions.
More precisely, we claim
that there exists a set~${\cal S}_v$ of simple regions such that
\begin{enumerate}
\item every~$S\in {\cal S}_v$ is a simple region between~$v$ and 
some vertex in~$\None^\ast(v)$, 
\item $\Ntwo^\ast(v) \subseteq \bigcup_{S \in {\cal S}_v}V(S)$, and
\item $|{\cal S}_v| \le 2\cdot |\None^\ast(v)|$.
\end{enumerate}

The idea for the construction of the set~${\cal S}_v$
is similar to the greedy-like 
construction of a maximal region decomposition (see Fig.~\ref{fig:regdec}).
Starting with~${\cal S}_v$ as empty set, one iteratively adds
a {\em simple} region~$S(v,x)$ between~$v$ and some vertex~$x \in \None^\ast(v)$
to the set~${\cal S}_v$ in such a way that (1) ${\cal S}_v \cup \{S(v,x)\}$ 
contains more~$\Ntwo^\ast(v)$-vertices than ${\cal S}_v$, 
(2) $S(v,x)$ does not cross any region in ${\cal S}_v$ and
(3) $S(v,x)$ is maximal (in space) under all simple regions~$S$
between $v$ and $x$ that do not cross any region in ${\cal S}_v$.
The fact that we end up with at most $2\cdot|\None^\ast(v)|$ many regions
can be proven by an argument on the induced graph~$G_{{\cal S}_v}$. 

Since, by Lemma~\ref{lem:simpleregion}, every simple
region~$S(v,x)$ with $x \in \None^\ast(v)$ contains 
at most seven vertices---not counting the vertices~$v$ and~$x$
which clearly cannot be in~$\Ntwo^\ast(v)$---we conclude that
$|\Ntwo^\ast(v)| \le 7\cdot |{\cal S}_v| \le 14 \cdot |\None^\ast(v)| \le
28\cdot \deg_{G_{\cal R}}(v)$.
%
{F}rom the fact that $V\setminus V({\cal R}) \subseteq D \cup (\bigcup_{v \in D} \Nthree(v)) \cup (\bigcup_{v \in D} \Ntwo^\ast(v))$ (see above)
we then get 
$$
|V\setminus V({\cal R})| \le |D| + |D| + \sum_{v \in D} |\Ntwo^\ast(v)| \le 2\cdot|D| + 28 \sum_{v \in D} \deg_{G_{\cal R}}(v) \le 2\cdot|D|+ 56 \cdot |{\cal R}|.
$$
\end{proof}

\begin{figure}[t]
\begin{center}
\begin{picture}(0,0)%
\epsfig{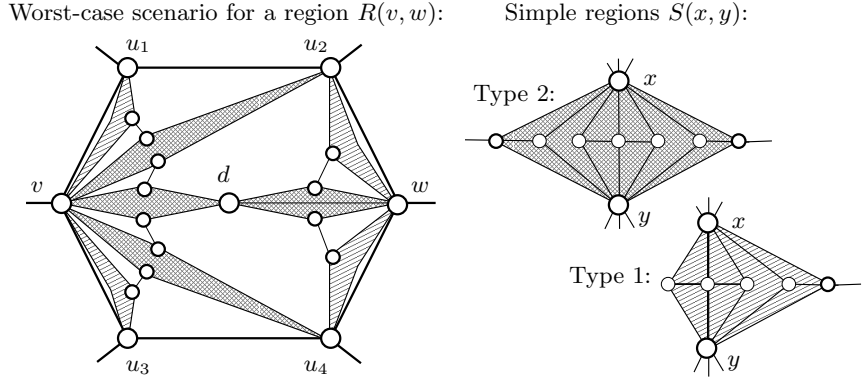}%
\end{picture}%
\setlength{\unitlength}{1865sp}%
\begingroup\makeatletter\ifx\SetFigFont\undefined%
\gdef\SetFigFont#1#2#3#4#5{%
  \reset@font\fontsize{#1}{#2pt}%
  \fontfamily{#3}\fontseries{#4}\fontshape{#5}%
  \selectfont}%
\fi\endgroup%
\begin{picture}(11332,5023)(2671,-5408)
\put(10126,-4111){\makebox(0,0)[lb]{\smash{\SetFigFont{9}{10.8}{\rmdefault}{\mddefault}{\updefault}Type 1:}}}
\put(8851,-1681){\makebox(0,0)[lb]{\smash{\SetFigFont{9}{10.8}{\rmdefault}{\mddefault}{\updefault}Type 2:}}}
\put(2671,-601){\makebox(0,0)[lb]{\smash{\SetFigFont{9}{10.8}{\rmdefault}{\mddefault}{\updefault}Worst-case scenario for a region $R(v,w)$:}}}
\put(9271,-601){\makebox(0,0)[lb]{\smash{\SetFigFont{9}{10.8}{\rmdefault}{\mddefault}{\updefault}Simple regions $S(x,y)$:}}}
\put(12226,-5206){\makebox(0,0)[lb]{\smash{\SetFigFont{9}{10.8}{\rmdefault}{\mddefault}{\updefault}$y$}}}
\put(8026,-2896){\makebox(0,0)[lb]{\smash{\SetFigFont{9}{10.8}{\rmdefault}{\mddefault}{\updefault}$w$}}}
\put(5446,-2761){\makebox(0,0)[lb]{\smash{\SetFigFont{9}{10.8}{\rmdefault}{\mddefault}{\updefault}$d$}}}
\put(4231,-5281){\makebox(0,0)[lb]{\smash{\SetFigFont{9}{10.8}{\rmdefault}{\mddefault}{\updefault}$u_3$}}}
\put(6931,-5281){\makebox(0,0)[rb]{\smash{\SetFigFont{9}{10.8}{\rmdefault}{\mddefault}{\updefault}$u_4$}}}
\put(6931,-1006){\makebox(0,0)[rb]{\smash{\SetFigFont{9}{10.8}{\rmdefault}{\mddefault}{\updefault}$u_2$}}}
\put(2971,-2896){\makebox(0,0)[lb]{\smash{\SetFigFont{9}{10.8}{\rmdefault}{\mddefault}{\updefault}$v$}}}
\put(4231,-1006){\makebox(0,0)[lb]{\smash{\SetFigFont{9}{10.8}{\rmdefault}{\mddefault}{\updefault}$u_1$}}}
\put(12286,-3376){\makebox(0,0)[lb]{\smash{\SetFigFont{9}{10.8}{\rmdefault}{\mddefault}{\updefault}$x$}}}
\put(11041,-3301){\makebox(0,0)[lb]{\smash{\SetFigFont{9}{10.8}{\rmdefault}{\mddefault}{\updefault}$y$}}}
\put(11116,-1471){\makebox(0,0)[lb]{\smash{\SetFigFont{9}{10.8}{\rmdefault}{\mddefault}{\updefault}$x$}}}
\end{picture}%

\end{center}     
\caption{\label{fig:bigregion} 
The left-hand diagram shows a worst-case scenario for 
a region~$R(v,w)$ between two vertices~$v$ and~$w$ in a reduced planar graph
(cf. the proof of Proposition~\ref{prop:region}).
Such a region may contain up to four vertices from~$\None(v,w)$,
namely~$u_1,u_2,u_3$, and~$u_4$. 
The vertices from~$R(v,w)$ which belong to the sets~$\Ntwo(v,w)$
and~$\Nthree(v,w)$ can be grouped into so-called simple regions
of Type 1 (marked with a line-pattern) or of Type 2 (marked with 
a crossing-pattern); the structure of such simple regions~$S(x,y)$ is
given in the right-hand part of the diagram.
In~$R(v,w)$ there might be two simple regions~$S(d,v)$ and~$S(d,w)$ (of Type 2), 
containing vertices from~$\Nthree(v,w)$. 
And, we can have up to six simple regions of vertices from~$\Ntwo(v,w)$:
$S(u_1,v),S(v,u_3),S(u_4,w),S(w,u_2),S(u_2,v)$, 
and $S(u_4,v)$ (among these, the
latter two can be of Type 2 and the others are of Type~1).
See the proof of Proposition~\ref{prop:region} for details.}

\end{figure}

We now investigate the maximal size of a region in a 
reduced graph. The worst-case scenario for a region in 
a reduced graph is depicted in Fig.~\ref{fig:bigregion}.
\begin{proposition}
\label{prop:region}
A region~$R$ of a plane reduced graph contains at most $55$~vertices, i.e.,
$|V(R)| \le 55$. 
\end{proposition}
\begin{proof}
Let~$R=R(v,w)$ be a region between vertices $v,w \in V$. 
As in the proof of Lemma~\ref{lem:simpleregion}, we count the number
of vertices in~$V(R)\subseteq N[v,w]$ which belong to~$\None(v,w)$, $\Ntwo(v,w)$,
and $\Nthree(v,w)$, separately.

We start with the number of vertices in~$\Nthree(v,w) \cap V(R)$.
Since the graph is assumed to be reduced, by Remark~\ref{rem:rule1and2},
we know that all vertices in~$\Nthree(v,w)$ need to be dominated
by a single vertex from~$\Ntwo(v,w) \cup \Nthree(v,w)$. Denote by~$d$ the
vertex which dominates all vertices in~$\Nthree(v,w)$.
Since all vertices in~$\Nthree(v,w)$ are also dominated by
$v$ or~$w$, we may write~$\Nthree(v,w) = S(d,v) \cup S(d,w)$ where
$S(d,v) \subseteq N(d) \cap N(v)$ and $S(d,w) \subseteq N(d) \cap N(w)$.
In this way, $S(d,v)$ and $S(d,w)$ form simple regions between
$d$ and~$v$, and $d$ and~$w$, respectively. 
In Fig.~\ref{fig:bigregion} these simple regions~$S(d,v)$ 
and~$S(d,w)$ (of Type 2) are drawn with a crossing pattern. 
By Lemma~\ref{lem:simpleregion}
we know that~$S(d,v)$ and $S(d,w)$ both contain at most seven vertices each, not
counting the vertices $d$,~$v$ and $d$,~$w$, respectively.
Since~$d$ maybe from~$\Nthree(v,w)$, we obtain $|\Nthree(v,w)\cap V(R)| \le 2\cdot7 + 1 = 15$.

It is clear that vertices in~$\None(v,w) \cap V(R)$ need to be 
on the boundary of~$R$, since, by definition of~$\None(v,w)$, 
they have a neighbor outside~$N(v,w)$.
The region~$R$ is enclosed by two paths~$P_1$ and~$P_2$ between
$v$ and~$w$ of length at most three each. 
Hence, there can be at most four vertices in~$\None(v,w) \cap V(R)$, 
where this worst-case holds if
$P_1$ and~$P_2$ are disjoint and have length exactly three each.
Consider Fig.~\ref{fig:bigregion}, which shows a region enclosed
by two such paths. Suppose that the four vertices on the boundary besides
$v$ and~$w$ are~$u_1,u_2,u_3$, and $u_4$.

Finally, we count the number of vertices in~$\Ntwo(v,w) \cap V(R)$.
It is important to note that, by definition of~$\Ntwo(v,w)$, every 
such vertex needs to have a neighbor in~$\None(v,w)$ and at the same
time needs to be a neighbor of either~$v$ or~$w$ (or both).
Hence, $\Ntwo(v,w) = \bigcup_{i=1}^4 (S(u_i,v) \cup S(u_i,w))$,
where $S(u_i,v) \subseteq N(u_i) \cap N(v)$ and $S(u_i,w) \subseteq N(u_i) \cap N(w)$.
All the sets $S(u_i,v)$ and~$S(u_i,w)$, where $1 \le i \le 4$, form 
simple regions inside~$R$. Due to planarity, however, there cannot exist 
all eight of these regions. In fact, in order to avoid crossings, the worst-case
scenario is depicted in Fig.~\ref{fig:bigregion} where six of these
simple regions exist (they are drawn with a line-pattern in
the figure).\footnote{%
Observe that regions $S(u_1,w)$ and $S(u_3,w)$ would cross
the regions~$S(u_2,v)$ and~$S(u_4,v)$, respectively.}
Concerning the type of these simple regions, it is not hard to verify,
that in the worst-case there can be two among these six regions of Type~2, 
the other four of them being of Type~1. 
In Fig.~\ref{fig:bigregion}, the simple regions~$S(u_2,v)$ and $S(u_4,v)$
are of Type~2 (having two connections to vertices outside the
simple region), and the simple regions~$S(u_1,v),S(u_2,w),S(u_3,v)$,
and $S(u_4,w)$  are of Type 1 
(having only one connection to vertices outside the region; a second
connection to vertices outside the region is not possible because of the 
edges~$\{u_1,v\},\{u_2,w\},\{u_3,v\}$, and~$\{u_4,w\}$).
In summary, the worst-case number of vertices in $\Ntwo(v,w) \cap V(R)$ 
is given by four times the number of vertices of a simple region 
of Type 1 and two times the number of vertices of a simple region
of Type 2; each time, of course, excluding vertices from~$\{u_1,u_2,u_3,u_4,v,w\}$.
By Lemma~\ref{lem:simpleregion} this amounts to 
$|\Ntwo(v,w) \cap V(R)| \le 4\cdot(3+2\cdot 1) + 2\cdot (3+2 \cdot2) = 34$.\footnote{%
Note that for the size of, e.g., a region~$S(u_i,v)$ we do not have to count~$u_i$
and~$v$, since they are not vertices in~$\Ntwo(v,w)$.}

The claim now follows from the fact that $V(R) = \{v,w\} \cup (V(R) \cap \Nthree(v,w))
\cup (V(R) \cap \None(v,w)) \cup (V(R) \cap \Ntwo(v,w))$, which yields
$|V(R)| = 2 + 15 + 4 + 34 = 55$.
\end{proof}

In summary, in order to prove Theorem~\ref{thm}
we first of all observe that, for a graph~$G$ with minimum dominating set~$D$, 
by Proposition~\ref{prop:dec} and Lemma~\ref{lem:thin},
we can find a $D$-region decomposition~$\cal R$ of~$G$ with at most
$3\gamma(G)$~regions, i.e., $|{\cal R}| \le 3 \gamma(G)$. 
By Proposition~\ref{prop:region}, we know that 
$
|V({\cal R}) | \le \sum_{R \in \cal R} |V(R)| \le 55|{\cal R}|. 
$
By Proposition~\ref{prop:leftovers}, we have
$|V \setminus V({\cal R}) | \le  2|D|+ 56|{\cal R}|$. 
Hence, we get $|V|  \le 2|D| + 111|{\cal R}| \le 335 \, \gamma(G)$.


\section{Concluding Remarks and Experimental Results}
\label{sec:concl}
In this work, two lines of research meet.
On the one hand, there is {\sc Dominating Set},
one of the NP-complete core problems of combinatorial 
optimization and graph theory. On the other hand, the second line of research is that 
of algorithm engineering and, in particular, the power of data 
reduction by efficient preprocessing.
Presenting two simple and easy to implement reduction rules 
for {\sc Dominating Set}, we proved that for planar graphs 
a linear size problem kernel can be efficiently constructed.
Our result complements and partially improves 
previous results~\cite{ABFN00,AFFFNRS01,AFN01a,AFN01b}
on the parameterized complexity of {\sc Dominating Set} on 
planar graphs. 
We emphasize that the proven bound on the problem kernel
size is a pure worst-case upper bound. In practice, we obtained
much smaller problem kernels (see below).

An immediate open question is to further 
lower the worst-case upper bound on the size of the problem kernel,
improving the constant factor to values say around~10.
This would bring the problem kernel for {\sc Dominating Set}
on planar graphs into ``dimensions'' as known for {\sc Vertex Cover},
where it is of ``optimal'' size~$2k$~\cite{CKJ01}.
This could be done by either improving the analysis given 
or (more importantly) further improving the given reduction rules
or both. Improving the rules might be done by further extending 
the concept of neighborhood to more than two vertices.
{F}rom a practical point of view, however, one also has to take into account 
to keep the reduction rules as simple as possible in order to avoid 
inefficiency due to increased overhead.
It might well be the case that additional, more complicated 
reduction rules only improve the worst case bounds, but are of little or no
practical use due to their computational overhead.

It might be interesting to see whether similar reduction rules
with a provable guarantee on the size of the reduced instances
can also be found for variations of {\sc dominating set} problem,
such as {\sc total dominating set}, or {\sc perfect dominating set} 
(see~\cite{Tel94} for a description of such variants).

Finally, we mention that the techniques in this paper are of a topological nature 
and can be carried over to prove a similar result for {\sc dominating set} on 
graphs of bounded genus. An open question is whether a linear problem kernel 
can also be proven for other graph classes such as, e.g., disk intersection graphs, 
for which the parameterized complexity of {\sc dominating set} is not known (see~\cite{AF02}).

\medskip
\noindent
{\bf Experimental studies.}
We briefly report on 
the efficiency of the given reduction rules
in practice. The performance of the preprocessing was measured
on a set of combinatorial random planar graphs of various sizes. More precisely,
we created eight sample sets of 100 random planar graphs each, containing
instances with 
100, 500, 750, 1000, 1500, 2000, 3000, and 4000 vertices.
%
%
%
The preprocessing seems,
at least on the given random sample sets, to be very effective.
As a general rule of thumb, we may say that, in all of the cases, 
\begin{itemize}
\item more than~$79\%$ of the vertices and
\item more than~$88\%$ of the edges
\end{itemize}
were removed from the graph. 
Moreover, the reduction rules determined a very high percentage 
(for all cases approximately~$89\%$) of the vertices of an optimal 
dominating set. 
The overall running time for the reduction ranged from
less than one second (for small graph instances with 100 vertices)
to around 30 seconds (for larger graph instances with 4000 vertices).

We remark that, in our experiments,  
we used a slight modification of the reduction rules:
Formally, when Rule~\ref{rule1} or Rule~\ref{rule2} is applied and
some vertex~$v$ is determined to belong to an optimal dominating,
the reduction rules attach a gadget vertex~$v'$ of degree one to~$v$.
In our setting, we simply removed the vertex~$v$ from the graph and 
``marked'' its neighbors as being already dominated. In this sense, we
dealt with an annotated version of {\sc dominating set}, where the
input instances are black-and-white graphs consisting of two types of vertices: 
black vertices which still need to be dominated; and white vertices which are
assumed to be already dominated. 
A slight modification makes Rule~\ref{rule1} and Rule~\ref{rule2} applicable
to such instances as well.

Finally, we enriched our reduction rules by further heuristics. 
We additionally used three (very simple) extra rules
that were presented in the search tree algorithm in~\cite{AFFFNRS01}. These extra rules 
are concerned with the removal of white vertices in such black-and-white graphs for
the annotated version of {\sc dominating set} (for the details and
their correctness see~\cite{AFFFNRS01}): (1) delete a white
vertex of degree zero or one; (2) delete a white vertex of degree two if its
neighbors are at distance at most two from each other; (3) delete a white vertex of degree three
if the subgraph induced by its neighbors is connected.

Enriching our reduction rules with these extra rules led to 
a very powerful data reduction on our set of random instances described above.
We observed that in this extended setting, the running times for the data
reduction went down to less than half a second (for graphs of 100 vertices) and
less than eight seconds (for graphs of 4000 vertices) in average. Most interestingly, 
the combination of these rules removed, in average, 
\begin{itemize}
\item more than~$99.7\%$ of the vertices and
\item more than~$99.8\%$ of the edges
\end{itemize}
of the original graph. A similarly high perentage of the vertices that belong to an optimal dominating
set could be detected.

%
%
Finally, in future work it will be our special concern 
to further extend our experimental studies 
to meaningful real-world input instances and/or non-planar input graphs.

\medskip

\begin{small}
\noindent {\bf Acknowledgements.}
For two years, besides ourselves the linear size problem
kernel question for {\sc Dominating Set} on planar graphs
has taken the attention of numerous people, all of whom we owe 
sincere thanks for their insightful and inspiring remarks and ideas.
Among these people we particularly would like to mention
Frederic Dorn, Henning Fernau, Jens Gramm, Michael Kaufmann, Ton Kloks, Klaus Reinhardt, 
Fran Rosamond, Peter Rossmanith, Ulrike Stege, and Pascal Tesson.
Special thanks go to Henning for the many hours he spent with us 
on ``diamond discussions'' and for pointing us to a small error concerning
the application of the linear problem kernel and to Frederic for 
again doing a perfect implementation job, which also
uncovered a small error in a previous version of reduction Rule~2.
\end{small}

\bibliographystyle{plain}


\end{document}